# First time calculation of the depletion region width and barrier capacitance of practical diffused semiconductor junctions

**Miron J. Cristea**

„Politehnica" University of Bucharest 061134, Bucharest, Romania

E-mail: mcris@lydo.org Tel: 40-721-245-673 Fax: 40-214-300-555

**Abstract**

Based on semiconductor materials fundamental equations, the calculation of the depletion region width and barrier capacitance of practical diffused -Gaussian profile- semiconductor junctions was achieved for the first time in this work. The obtained formulas are valid for p-n junctions, Schottky junctions, hetero-junctions and other types of semiconductor junctions.

**1. Introduction**

The Gauss's law for a semiconductor junction gives in the one-dimensional case [1]:

$$\frac{dE}{dx} = \frac{\rho(x)}{\varepsilon} \qquad (1)$$

where $E$ is the electric field across the junction, $\varepsilon$ represents the permittivity of the semiconductor material and $\rho$ is the space charge density within the space charge region (SCR)[2,3]:

$$\rho \cong q(N_d - N_a) \qquad (2)$$

$N_d$ being the donor doping concentration, $N_a$ the acceptor doping concentration and $q$ the elementary charge. Depletion approximation is considered.

The distribution of the electric potential $u$ across the junction is in accordance with [1,2]:

$$E = -\frac{du}{dx} \qquad (3)$$

Practical diffused junctions have a Gaussian profile of impurities [3],

$$N_d(x) = N_0 \exp\left(-\frac{x^2}{L_d^2}\right) \qquad (4)$$

where $N_0$ is the surface concentration and $L_d$ is the technological diffusion length

$$L_d = 2\sqrt{D_i \cdot t_d} \qquad (5)$$

with $D_i$ - the doping impurity diffusion constant at certain diffusion temperature and $t_d$ the diffusion time. Eq. (1) can not be integrated with Gaussian function (4). Because of that, by means of some mathematical transformations, an integral formula that allows the calculation of the depletion width of such junctions was derived (Appendix 1):

$$\int_{SCR} \frac{x\rho(x)}{\varepsilon}dx = V_{bi} - V_F \qquad (6)$$

where $V_{bi}$ is the built-in potential and $V_F$ is the external forward bias applied to the junction. In the case of a reverse-biased junction, $-V_F$ is replaced by $V_R$:

$$\int_{SCR} \frac{x\rho(x)}{\varepsilon}dx = V_{bi} + V_R \qquad (7)$$

where $V_R$ is the external reverse bias.

In the case of homogenous semiconductor junctions, the formula can be written as:

$$\frac{1}{\varepsilon} \int_{SCR} x\rho(x)dx = V_{bi} + V_R \qquad (8)$$

since the permittivity is constant throughout the material.

However, in the case of hetero-junctions or other types of junctions in which more than one material is encountered, the following form of equation (7) should be applied:

$$\int_{SCR1} \frac{x\rho(x)}{\varepsilon_1} dx + \int_{SCR2} \frac{x\rho(x)}{\varepsilon_2} dx + ... + \int_{SCRn} \frac{x\rho(x)}{\varepsilon_n} dx = V_{bi} + V_R \qquad (9)$$

where SCR1, SCR2... SCRn are the fractions of the overall space charge region corresponding to the *n* semiconductor materials used for the junction fabrication.

## 2. The derivation of the space charge region width:

In Figure 1 are depicted the distribution of doping impurities and the space charge region of a practical diffused p-n junction. The equation describing the distribution of the diffused profile of impurities in Fig.1 is (4).

Supposing that the junction is reverse biased, and that the semiconductor material is homogenous (e.g. silicon) equation (8) gives:

$$\frac{qN_0}{\varepsilon} \int_{SCR} x \exp\left(-\frac{x^2}{L_d^2}\right) dx \cong \frac{qN_0}{\varepsilon} \int_{x_j}^{W_{SC}+x_j} x \exp\left(-\frac{x^2}{L_d^2}\right) dx = V_R + V_{bi} \qquad (10)$$

since the extension of the SCR in the heavily doped side of the junction is negligible. The integration of (10) leads to:

$$\frac{qN_0 L_d^2}{2\varepsilon} \left[ \exp\left(-\frac{x_j^2}{L_d^2}\right) - \exp\left(-\frac{(W_{SC}+x_j)^2}{L_d^2}\right) \right] = V_R + V_{bi} \qquad (11)$$

From (11) the space charge region width is obtained:

$$W_{SC} = L_d \sqrt{\ln \frac{1}{\exp\left(-\frac{x_j^2}{L_d^2}\right) - \frac{2\varepsilon}{qN_0 L_d^2}(V_R + V_{bi})}} - x_j \quad (12)$$

In the case of a slightly forward-biased junction, $V_R$ is to be replaced by $-V_F$.

## 3. The barrier capacitance

The specific capacitance per area unit is given by [2]:

$$C_b = \frac{\varepsilon}{W_{SC}} \quad (13)$$

Using (12), the next formula is obtained:

$$C_b = \frac{\varepsilon}{L_d \sqrt{-\ln\left[\exp\left(-\frac{x_j^2}{L_d^2}\right) - \frac{2\varepsilon}{qN_0 L_d^2}(V_R + V_{bi})\right]} - x_j} \quad (14)$$

For shallow junctions, $x_j$ is negligible compared to $W_{SC}$, therefore the barrier capacitance can be written as:

$$C_b \cong \frac{\varepsilon}{L_d} \left\{ \sqrt{-\ln\left[\exp\left(-\frac{x_j^2}{L_d^2}\right) - \frac{2\varepsilon}{qN_0 L_d^2}(V_R + V_{bi})\right]} \right\}^{-1} \quad (15)$$

In addition, the SCR width for such junctions is:

$$W_{SC} = L_d \sqrt{-\ln\left[\exp\left(-\frac{x_j^2}{L_d^2}\right) - \frac{2\varepsilon}{qN_0 L_d^2}(V_R + V_{bi})\right]} \qquad (16)$$

## 4. The case of junctions with $x_j \ll L_d$ (deep diffused junctions)

For some junctions, the condition $x_j \ll L_d$ is fulfilled. These junctions can be called deep diffused junctions. For them, the exponential term in (15) and (16) can be taken as unity. The following simplified formulas are obtained:

$$W_{SC} = L_d \sqrt{-\ln\left[1 - \frac{2\varepsilon}{qN_0 L_d^2}(V_R + V_{bi})\right]} \qquad (17)$$

$$C_b \cong \frac{\varepsilon}{L_d}\left\{\sqrt{-\ln\left[1 - \frac{2\varepsilon}{qN_0 L_d^2}(V_R + V_{bi})\right]}\right\}^{-1} \qquad (18)$$

**Conclusion**

In this work, new formulas that give the depletion region width and barrier capacitance of practical diffused semiconductor junctions have been calculated for the first time. The particular cases of shallow and deep diffused junctions have been investigated also.

## Appendix 1

As mentioned, the Gauss's law gives:

$$\frac{dE}{dx} = \frac{\rho(x)}{\varepsilon} \quad (A1)$$

By writing (A1) as

$$dE = \frac{\rho(x)}{\varepsilon} dx \quad (A2)$$

multiplying this equation by $x$ and taking into account that

$$d(xE) = xdE + Edx \quad (A3)$$

the next is obtained:

$$d(xE) - Edx = \frac{\rho(x)}{\varepsilon} xdx \quad (A4)$$

The integration of Eq. (A4) over the space charge region (*SCR*) of the junction gives:

$$\int_{SCR} \frac{x\rho(x)}{\varepsilon} dx = \int_{SCR} d(xE) - \int_{SCR} Edx \quad (A5)$$

and taking (3) into account ($Edx = -du$):

$$\int_{SCR} \frac{x\rho(x)}{\varepsilon} dx = \int_{SCR} d(xE) + \int_{SCR} du \quad (A6)$$

Since the electric field is zero at both ends of the SCR, the first term in the right hand of Eq. (A6) vanishes, and the next equation is obtained:

$$\int_{SCR} \frac{x\rho(x)}{\varepsilon} dx = V_{bi} - V_F \quad (A7)$$

where $V_{bi}$ is the junction built-in voltage and $V_F$ is the external forward bias applied to the junction.

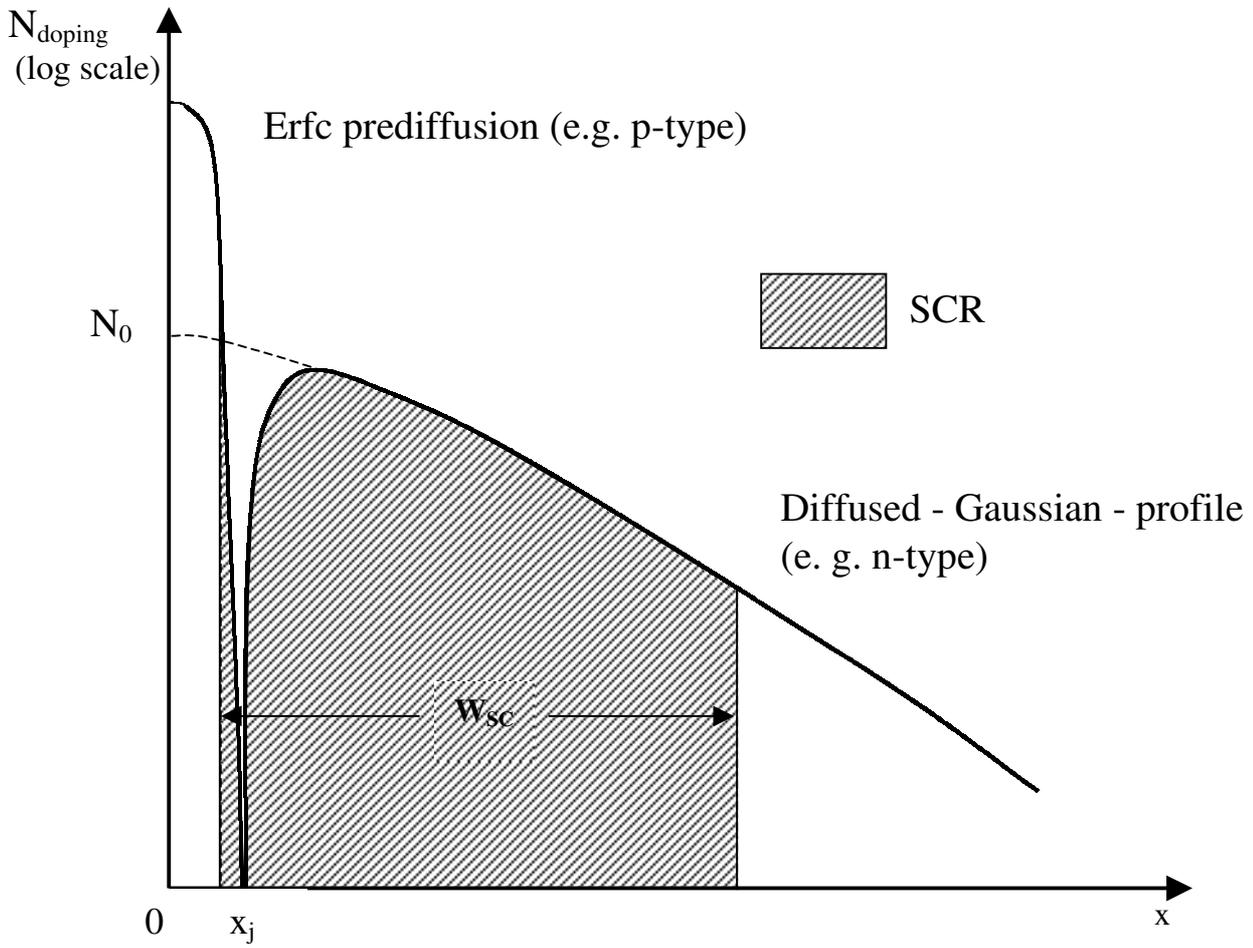

**Figure 1**. Doping profiles and space charge region (SCR) in a practical semiconductor p-n junction